\documentclass[a4paper]{article}

\usepackage[english]{babel}
\usepackage[utf8x]{inputenc}
\usepackage[T1]{fontenc}
\usepackage{authblk}

\usepackage[a4paper,top=3cm,bottom=2cm,left=3cm,right=3cm,marginparwidth=1.75cm]{geometry}

\usepackage{amsmath}
\usepackage{graphicx}
\usepackage[colorinlistoftodos]{todonotes}
\usepackage[colorlinks=true, allcolors=blue]{hyperref}
\usepackage{graphicx}
\usepackage{subcaption}

\usepackage{xcolor}

\begin{document}

\title{Life in the coffee-ring: how evaporation-driven density gradients dictate the outcome of inter-bacterial competition}

\author[1]{David Yanni}
\author[1]{Arben Kalziqi}
\author[2]{Jacob Thomas}
\author[2]{Siu Lung Ng}
\author[1]{Skanda Vivek}
\author[2]{William C. Ratcliff}
\author[2]{Brian K. Hammer}
\author[1]{Peter J. Yunker}
\affil[1]{Georgia Tech School of Physics}
\affil[2]{Georgia Tech School of Biological Sciences}
\date{}
\maketitle
\begin{abstract} When a drop dries, it often leaves a ring-shaped stain through a ubiquitous phenomenon known as the coffee-ring effect. This also occurs when the liquid contains suspended microbes; evaporation leaves cells at higher concentrations in the ring than the drop interior. Using biofilm experiments and cellular automata simulations, we show that the physical structure created by the coffee-ring effect can dramatically alter the outcome of inter-bacterial competition. We experimentally study this effect using two strains of \textit{Vibrio cholerae} that compete using a contact-dependent killing mechanism termed the type 6 secretion system. By creating a heterogeneous density profile, the coffee-ring effect changes the outcome of competition: the bacterial strain that wins in the low-density interior loses in the far denser ring. Through simulations parameterized with experimentally-determined density profiles, we recapitulate our experimental findings. We examine the role of a \textit{V. cholerae} strain's frequency, its relative efficacy at killing competitors, and the initial concentration of cells in the droplet in determining the outcome of competition. By scaling from individual cellular interactions to overall changes in strain frequency, our work demonstrates how the coffee-ring effect plays a powerful role in structuring microbial communities, indirectly driving ecological changes in community composition.
\end{abstract}

\section{Introduction}
Bacteria often form dense, surface-attached colonies known as biofilms \cite{Review}. Understanding how these complex microbial communities are structured is of fundamental and practical interest, as they dramatically change evolutionary outcomes \cite{Tolker-Nielson}\cite{Nadell}\cite{Nielson}\cite{density_and_spatial_B_subtilits} and underlie recalcitrant medical \cite{review infectious 1}\cite{review infectious 2} and industrial \cite{food industry}\cite{mining industry} problems. Biofilms are highly plastic and can be formed from a few founder cells and remain relatively clonal, or they can arise from whole populations comprising diverse microbial taxa\cite{Davey Review}. While a great deal of effort has gone into understanding how biofilms develop, far less attention has been paid to understanding how common physical forces structure their early assembly \cite{MechWorldBact}.

When a drop dries, it almost always leaves behind a ring-shaped stain. A gradient in evaporation rate - lowest at the drop center, highest at the drop edge - drives a fluid flow from the center to the edge, carrying anything in the drop to the periphery \cite{Deegan}\cite{Deegan2}\cite{Deegan3}. When evaporation finishes, it leaves behind a deposit that is denser at the edge than in the interior. This phenomenon, known as the coffee-ring effect\cite{Deegan}\cite{Deegan2}\cite{Deegan3}\cite{Yunker_coffee_ring}, creates ring-shaped deposits of anything dissolved or suspended in a drop, from the oils in a cup of coffee \cite{Deegan}, to colloidal particles \cite{Deegan}\cite{Deegan2}\cite{Deegan3}\cite{Yunker_coffee_ring}, to cells\cite{coffee_ring_cells} \cite{nannochrom_coffee_ring}\cite{Thokchom}. Indeed, a drop of rain, fresh water, or a sneeze aerosol can contain millions of microbial cells\cite{millions_1}. When these drops dry on a surface, the coffee ring effect will play a central role in spatially-structuring the population. Along with these natural occurrences, this effect is likely present in most laboratory experiments wherein agar slabs are inoculated with small drops of liquid culture (Figure 1). Cells deposited in the densely packed ring immediately interact with each other, while cells in the dilute interior initially undergo clonal outgrowth. Thus, the coffee-ring-determined spatial variance of initial cell concentration may play a large role in mediating the local interactions between deposited cells, directly impacting fitness and biofilm structure.  

In this paper, we use a model system of interbacterial competition to examine consequences of spatial structuring via the the coffee-ring effect. Using simulations and experiments, we compete two strains of \textit{Vibrio cholerae} that can each kill the other (but not siblings) on contact via the type VI secretion system (T6SS). The T6SS is an apparatus $\sim$25\% of all sequenced gram negative bacteria use to translocate toxic effector proteins into the cytoplasm of a target cell on contact \cite{T6SS}, which can lead to clonal phase separation in biofilms\cite{McNally}. Interestingly, the strain that is superior in the low-density interior of the biofilm appears inferior in the coffee ring. We show that this context dependence is due to the coffee-ring effect, which generates a density heterogeneity between the coffee-ring and its interior. Further, we show that when competing two strains, three parameters must all be considered: the ratio of T6SS-mediated killing effectivenesses, the initial number ratio, and the initial concentration. Due to its ubiquity in both natural and laboratory settings, understanding the role of the coffee-ring effect in structuring microbial biofilms is of central biological importance. 

\section{Methods}
\subsection{Experiments}
The \textit{V. cholerae} strains used were derivatives of isolates C6706 and 692-79 \cite{Bernardy}, and were constructed as described in McNally, \textit{et al.} \cite{McNally}. The C6706 strain was tagged with an mKO fluorescent protein and is shown here in red. The 692-79 strain was labeled with mTFP1 and is shown here in blue. Cells were grown at $30^{\circ}$C in Lysogeny Broth (LB) liquid medium to an optical density (OD) $\sim 1$  and subsequently inoculated as $1\,\mu L$ drops onto LB agar (1.5\%) pads on glass slides. Imaging was performed using laser fluorescence confocal microscopy with a Nikon A1R, as described in \cite{McNally}. Agar pads were inoculated with a mixture of the two strains at a number ratio of 1:4 (C6706:692-79); this number ratio was chosen as it produces, after 24 hours of incubation, a nearly 50:50 final mixture of the two strains in the center of the biofilm. Suspensions of the two strains with the chosen number ratio were prepared immediately prior to inoculation by mixing different volumes of single-strain cultures grown to identical optical density. For a number ratio of 1:4, this meant mixing $1\,\mu L$ of liquid culture C6706 ($OD=1$) into $4\, \mu L$ of liquid culture 692-79 ($OD=1$) just before inoculation. 

\subsection{Simulations}

Our simulation is a 2D stochastic cellular automata model of biofilm member cells. Sites are arranged on a square lattice of varying length, from 400 (to generate the data in figure 2) to 2500 (to generate figure 1 C). Each site is occupied by a red cell, a blue cell, or an empty space (black). A cell can interact with any of its eight neighbors. A red (blue) cell may kill a neighboring blue (red) cell with probability $p_{rk}$  ($p_{bk}$), leaving an empty site; a red (blue) cell may double-into a neighboring empty site with probability $p_{rd}$ ($p_{bd}$). A simulated timestep is defined as choosing a site at random from the lattice, subjecting it to the transition rules, and repeating for either $400^2$ and $2500^2$  iterations (depending on the lattice size), so that each automaton is expected to have had the chance to reproduce (space permitting) once per timestep. \newline

\section{Results and Discussion}

Immediately after inoculation, we imaged biofilms using brightfield and fluorescence microscopy (Figure 1A and 1B). Indeed, cells deposit more densely in the coffee ring (Figure 1A, and 1B, left). After 24 hours of growth, biofilms composed of two mutually killing strains reproducibly exhibit three regimes of cellular arrangement (Figure 1 B, right. See SI Figure 1 for replicates). In the interior of the biofilm, T6SS mediated warfare causes clonal phase separation, as described in \cite{McNally}. On the biofilm periphery, outside of the blue annulus, cells undergo range expansion into uncolonized territory\cite{Hallatschek} \cite{Weinstein}. Surprisingly, while the frequency of the red strain with superior T6SS weaponry ($2$-fold better killing as measured by the number of surviving \textit{E. coli} cells in a killing assay \cite{Bernardy}) increases throughout the competition in the center, it actually decreases in the coffee-ring, leaving a blue ring. \newline

\begin{figure}[!ht]
\centering\includegraphics[width=\textwidth]{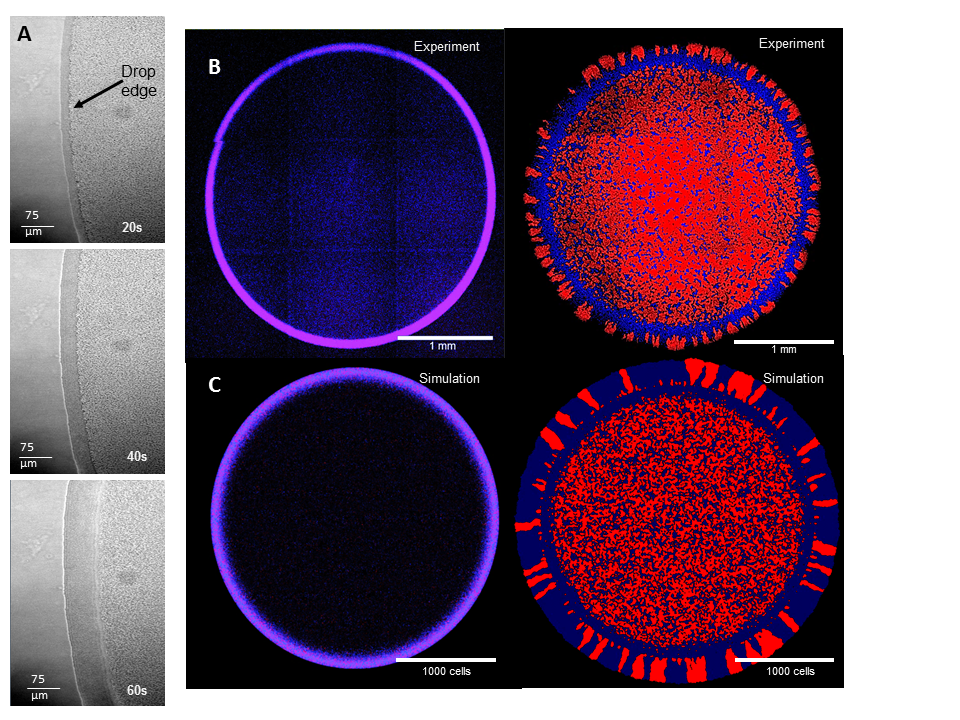}
\caption{ \textbf{A}: Series of images showing the build-up of cells at the periphery of a drying drop of liquid media on an agar substrate. i-iii: images taken at 20, 40,  and 60 seconds after inoculation, respectively. 
\textbf{B}: Left: Fluorescence image showing the distribution of the two strains 2 minutes after inoculation, just after the coffee ring has formed. Right: approximately 24 hours after inoculation.
\textbf{C}: Left: Initial condition of the simulation. Cells were seeded randomly, according to the measured initial concentration profile $x(r)$ in (\textbf{B}), at 4:1 blue to red, matching the initial distribution in (\textbf{B}) (see SI Figure 3). The total number of simulated cells is approximately commensurate with the experiment (\textbf{B}). Right: Simulated biofilm after 75 timesteps. Simulation parameters used were $ p_{bk}=0.77p_{rk}$ , and  $p_{bd}=0.95p_{rd}$.
}
\end{figure}

Simulations seeded with spatially uniform concentrations do not replicate the three observed regimes \cite{McNally} (SI Figure 2). However, a simulated biofilm seeded with the experimentally observed cellular concentration profile (Figure 1 B, SI Figure 3) had both the appearance and distinct morphological features of the experimental biofilm that had been incubated at $30^{\circ}$  C for 24 hours (Figure 1 C). \newline

Interestingly, simulations replicate these three regimes with biologically reasonable parameters. They "grow" for 75 time steps, allowing 75 opportunities to double; \textit{V. cholerae} cells divide every $\sim 20$ minutes in liquid media, so in a biofilm grown for 24 hours the maximum number of possible reproduction events per cell is consistent with our simulation value of 75. The number of automata we simulate is commensurate with the number of cells in a monolayer of biofilm ($\sim 7\times10^6$). With nearly matched growth rates between the two simulated strains ($p_{bd}=0.95\,p_{rd}$), and killing rates on the same order of magnitude ($ p_{bk}=0.77\,p_{rk}$), we capture key cellular behaviors with sensible parameters (see ref  \cite{Bernardy}).\newline 

To investigate how a strain with a lower growth rate, lower killing rate, or both, can drive its competitor to (or near) extinction in the coffee ring, we performed further simulations to probe how the following parameters impact competition:

\begin{center}
\begin{tabular}{c}
initial proportion of red cells: $r_{i}=\frac{\text{number of red cells}}{\text{number of red and blue cells}}$ \\ \\
killing rate ratio: $k_R = \frac{p_{rk}}{p_{bk}}$ \\ 
\\ 
initial concentration: $x_0=\frac{\text{number of sites occupied}}{\text{total number of sites}}$ 
\end{tabular}
\end{center}

We first investigated the effect of $r_i$ on competition outcomes. Simulations began with completely full square lattices randomly seeded with red and blue cells corresponding to different $r_{i}$ (see e.g. the blowup in Figure 2 A); after $t=100$, the final proportion of red cells depends logistically on the initial proportion (Figure 2 A). We tracked the proportion of red cells over time and declared the "winner" of the competition to be the strain which occupies more territory (which corresponds to cell number in a 2D lattice) at the end of the simulation. For example, for strains with equal killing and growth rates, the transition from one winner to the other occurs at a proportion of $r^*_{i}=0.5$ (Figure 2 A, rightmost curve with triangles). This is unsurprising, as whichever strain has even a slight initial numeric advantage is expected to win this competition.

Next, we investigated the role of the red strain's relative kill rate, $k_R$, on competition outcomes. We again started with completely full lattices, but this time varied $k_R$, along with $r_{i}$. Varying $k_R$ affects the transition proportion, $r^*_{i}$; increasing $k_R$ shifts the entire trend to the left without otherwise effecting qualitative changes (Figure 2 A).

Finally, we investigated the effect of the initial cellular concentration, $x_0$. The effect is readily visualized in Figure 2D, where a competition was initialized with cellular density decreasing from 100\% (left) to 0\% (right). Blue (an inferior killer, $k_R = 2$ that is initially more abundant $r_i = 0.25$) cells fare better in initially high concentration regimes. Conversely, low initial density favors the strain with the killing advantage, even when it is initially out-numbered by a significant margin. This effect is quantified in Figure 2 B. Here, we generate a family of logistic curves like the one shown in Figure 2 A by repeating all of the simulations as described above, but seeded with a particular initial concentration of cells, $x_0$. For each data point, we discern the minimum proportion of red cells needed to win the competition, $r_i^*$, for a given $k_R$ and $x_0$ (Figure 2 B). We find that large initial concentrations of cells favor the strain with a numeric advantage, even though it has a significantly inferior killing rate (like the blue strain in Figure 2 D), while small initial concentrations of cells favor the strain with a killing rate advantage.

The mechanism behind this counter-intuitive density dependence can be understood by considering the local environment a cell experiences at early times. We measured the average local fraction of red cells $\sim 1$ timestep after the start of each simulation, $<\widetilde{r}_i>$, and plotted the value of $<\widetilde{r}_i>$ when red cells win the competition as a function of $x_0$ and $k_R$. This collapses the curves of Figure 2B onto a single trend line. When a superior but outnumbered cell is deposited into a crowded area it is immediately confronted with its minority status, and any local frequency-dependent selection effects are likely to diminish its long term success. For example, for cells equipped with T6SS, even a strain that kills at twice the per-cell rate as its competitor is nevertheless at a competitive disadvantage if it immediately faces three times as many enemies within killing range. However, a superior but outnumbered cell that is deposited into a sparse area undergoes initially unimpeded clonal outgrowth and soon establishes a local majority wherein frequency dependent effects enhance its viability. Even if the global number ratio of red to blue cells is equivalent in a sparse or dense inoculate, differences in the local number ratio---ultimately the relevant parameter for local interactions----can effect different outcomes.

\begin{figure}[!h]
\begin{minipage}{\textwidth}
\centering
\includegraphics[width=\textwidth]{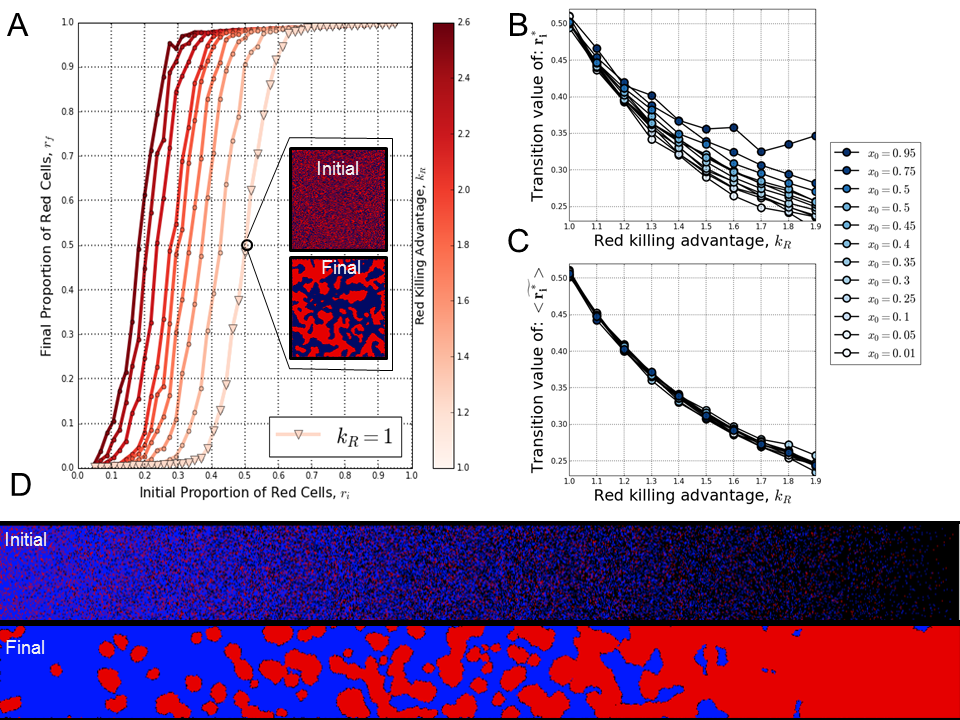}
\caption[blah]{
The final proportion of red cells, $r_{f}$, after a $100$ time step competition between red and blue cells in a square lattice of size $400^2$ cells, depends logistically on the initial proportion of red cells: $r_{f}=\frac{1}{1+e^{(-\beta(r_{i}-r_i^*))}}$.The blowup shows an example of the type of simulation from which $r_f$ is extracted, in this case $(r_i=0.5, r_f=0.5)$. $r_i^*$ can be understood as the initial proportion of red cells required to "win" the competition. The effect of varying killing rate disparity is plotted in \textbf{A} as solid lines representing the variation of $k_R$, from $k_R=1$ (far right curve with triangles) to $k_R=2.7$ (far left curve).  $r_i^*$ decreases as $k_R$ increases; it takes fewer red cells initially to win the competition if each red cell is more competitive.  The effect of varying the overall initial concentration of cells (of both types), $x_0$ is visualized in \textbf{D}, where a simulation was seeded with concentration varying from $1$ at the far left where all the sites are occupied by cells initially to $0$ at the far right where all the sites are initially empty. Note that, initially, blue outnumbers red 4:1 throughout the entire lattice ($r_i=.25$), but red is twice as effective at killing, ($k_R=2$). The state of the simulation after $100$ timesteps is shown on the bottom. \textbf{B} quantifies this concentration dependence. The darkest circles were generated from the family of curves shown in \textbf{A} (where the curves intersect $r_f=0.5$), and the rest of the data in \textbf{B} were generated similarly (see SI Figures 5 and 6). As $x_0$ decreases, the individually superior but outnumbered strain fares better. These data collapse if we plot against the variable $<\widetilde{r_i^*}>$, \textbf{C}. Whereas $r_i$ describes the global proportion of red to blue cells throughout the entire lattice at time zero, $\widetilde{r_i}(x,y)$ describes the initial$^\dagger$,  local$^\ddagger$ proportion of red to blue cells at a lattice site $(x,y)$. Taking a spatial average yields $<\widetilde{r_i}>$ , the transition value of which is  $<\widetilde{r_i^*}>$.\newline $^\dagger$ In fact $\widetilde{r_i}(x,y)$ is measured at early but non-zero time ($t\sim 1$ timesteps), see SI Figure 7. \newline $^\ddagger$ Here, by local we mean the nine-automaton 
neighborhood centered at $(x,y)$.}
\end{minipage}
\end{figure}

\section{Conclusions}
Drying droplets of water structure surface-attached microbial communities through the coffee-ring effect. Using a bacterial model system, we show how the deposition of \textit{V. cholerae} cells in the classic 'coffee ring' pattern changes competitive (and thus evolutionary) outcomes: T6SS-mediated killing is frequency-dependent, causing the superior killer to win in the low-density interior of the biofilm while losing in the high-density coffee ring. This has fundamental implications for understanding the spatial ecology of microbes, and allows for novel predictions: for example, our model predicts that the overall winner of the competition modeled in Figure 1 C can be flipped simply by changing the radius of the droplet (which changes the relative size of the interior relative to the ring). We anticipate the coffee-ring effect will be relevant to a broad swath of microbes that possess frequency and density dependent interactions (\textit{e.g.}, quorum sensing, or the production of extracellular metabolites and biofilm structural components). Given how difficult it is to avoid the coffee-ring effect when drying droplets of cells on agar media, it likely also plays a significant yet under-appreciated role in microbiological experiments.

\enlargethispage{20pt}

\section{Data Access}
All simulations were implemented in Python 3.5, using NumPy and SciPy for numerics, Matplotlib for plotting and the Python Imaging Library to generate images. Simulation source is available at: \newline \href{https://github.gatech.edu/dyanni3/V.Cholerae.Coffee.Ring/tree/master}{https://github.gatech.edu/dyanni3/V.Cholerae.Coffee.Ring/tree/master}

\section{Author Contributions}Designed and Interpreted Experiments:\newline
David Yanni, Arben Kalziqi, Jacob Thomas, Siu Lung Ng, William C. Ratcliff, Brian K. Hammer, Peter J. Yunker\newline
Helped with writing:\newline
David Yanni, 
Arben Kalziqi, 
Jacob Thomas, 
Siu Lung Ng, 
William C. Ratcliff, 
Brian K. Hammer, 
Peter J. Yunker\newline
Performed Imaging:\newline
David Yanni, 
Jacob Thomas\newline
Constructed and/or prepared V. cholerae strains:\newline	
Jacob Thomas, 
Siu Lung Ng, 
Brian K. Hammer\newline
Performed Simulations and Analysis:\newline
David Yanni

\section{Competing Interests}None of the submitted material has been published or is under consideration elsewhere. We do not have any related papers in press or under consideration, or any other competing interests.

\section{Funding Statement} This work was supported by National Science Foundation grants MCB-1149925 to B.H and IOS-1656549 to P.J.Y. and W.C.R. W.C.R. was also supported by a Packard Foundation Fellowship. P.Y. and W.C.R. were supported by the Georgia Tech Soft Matter Incubator.

\section{Acknowledgements}We would like to thank Bryan T. Weinstein for useful discussion.

\section{Supplementary Information}

\subsection{The \textit{V. cholerae} coffee-ring}

Replicates of the outcome of the competition between two strains of \textit{V. cholerae} exhibit similar final morphologies (supplemental figure 1). Additionally, color-swapped strains show a similar effect, further suggesting that production of fluorescent proteins does not affect their fitness (supplemental figure 1). Finally, defective-killer versions of these strains only shows a very faint coffee ring (supplemental figure 1).

The coffee-ring, constructed of \textit{V. cholerae} can clearly be seen in fluorescence (supplemental figure 3) and brightfield images (supplemental figure 4). Fluorescence microscopy also permits measurement of the local number ratio.

\subsection{Simulations}

Example images in supplemental figure 2 show the progression of simulations over time for five different sets of parameters. The data in supplemental figure 5 are extracted from simulations shown in supplemental figure 2. Then, a logistic fit, shown in supplemental figure 6, is applied, allowing the extraction of $r_{i}^{*}$.

Supplemental figure 7 demonstrates how simulations with the same number ratio between red and blue strains, $r$, can appear significantly different, and have very different local number ratios, $\widetilde{r_i}$. Histograms demonstrating the distribution of local configurations are shown as well.





\begin{figure}[!ht]
\centering
\includegraphics[width=\textwidth]{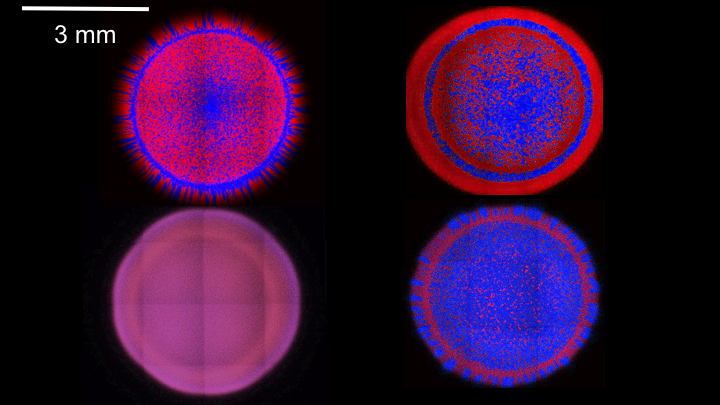}
\caption{\label{fig:cRings} Replicates of coffee-ring outcomes. The superior killer consistently wins in the interior of the drop. Note that these images have been recolored from red-green to red-blue to avoid issues with color-blindness. Bottom right sample contains color-swapped versions of the strains used in the main text. This serves as a control to show that the fluorescent proteins used have no discernible competitive effect. The biofilm shown on bottom left is composed of strains that cannot kill as in \cite{McNally} A faint coffee-ring is still visible but less pronounced. The number ratio of red to blue cells used here was the same as for biofilms composed of mutual killers, and the average clonal patch size is different in the coffee ring from the interior. There may also be differences in growth rates between these strains}
\end{figure}

\begin{figure}
\centering
\includegraphics[width=\textwidth]{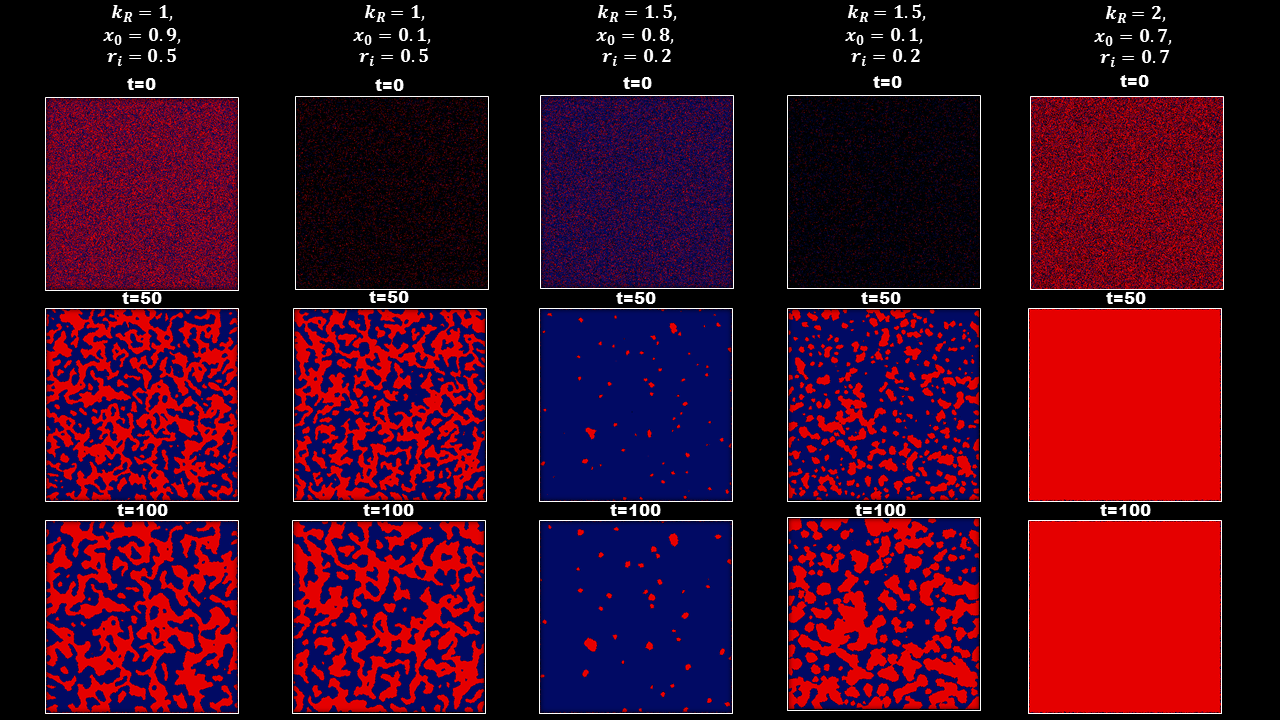}
\caption{\label{fig:square_competition} Examples of simulations used to generate data in main text figure 2. Each column represents the time-evolution of a single square biofilm. Simulation parameters are listed above the top image, including the initial proportion of red cells, $r_i$. As simulations were seeded with uniform initial cellular concentration they do not exhibit the coffee-ring or outgrowth features of our larger simulations or experiments. However, clonal phase separation as in \cite{McNally} does occur. To determine the outcome of these competitions the total number of red cells at $t=100$ is divided by the total number of cells at $t=100$. If this ratio, $r_f$, is greater than 0.5 then red is declared the competition winner. $r_f$ values from competitions such as these are plotted against $r_i$ values for different simulation parameters in main text figure 2A and SI figure 5}
\end{figure}

\begin{figure}[!ht]
\centering
\includegraphics[width=\textwidth]{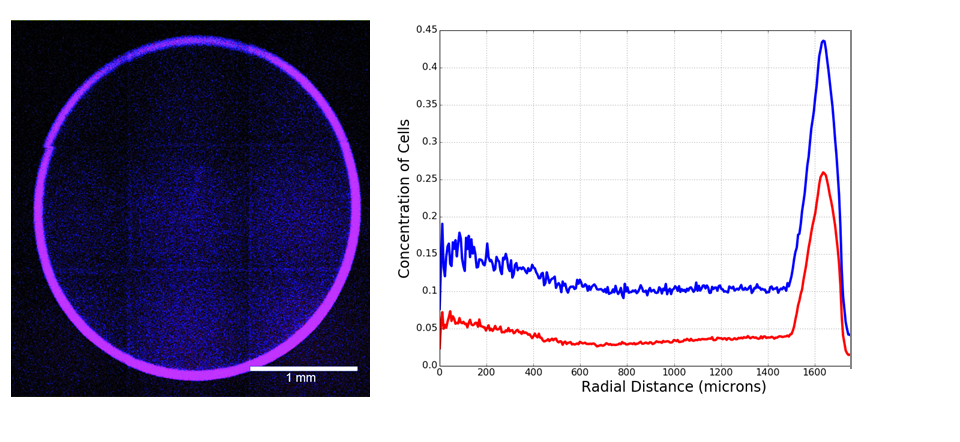}
\caption{\label{fig:rpr}Fluorescence microscopy images of a dried drop of liquid culture containing two strains of \textit{V. cholerae} show that the density is higher at the edge than in the middle (left). The initial radial-cellular-concentration profiles (right) are calculated from the image of the entire deposition. These profiles approximate the actual cellular deposit, and were used to seed the simulation pictured in the main text figure 1C. Here concentration is defined relative to maximum pixel intensity, which is taken to represent full area coverage by cells over the agar. Profiles were generated by azimuthally averaging over the pixels in the input image.}
\end{figure}

\begin{figure}[!ht]
\centering
\includegraphics[width=\textwidth]{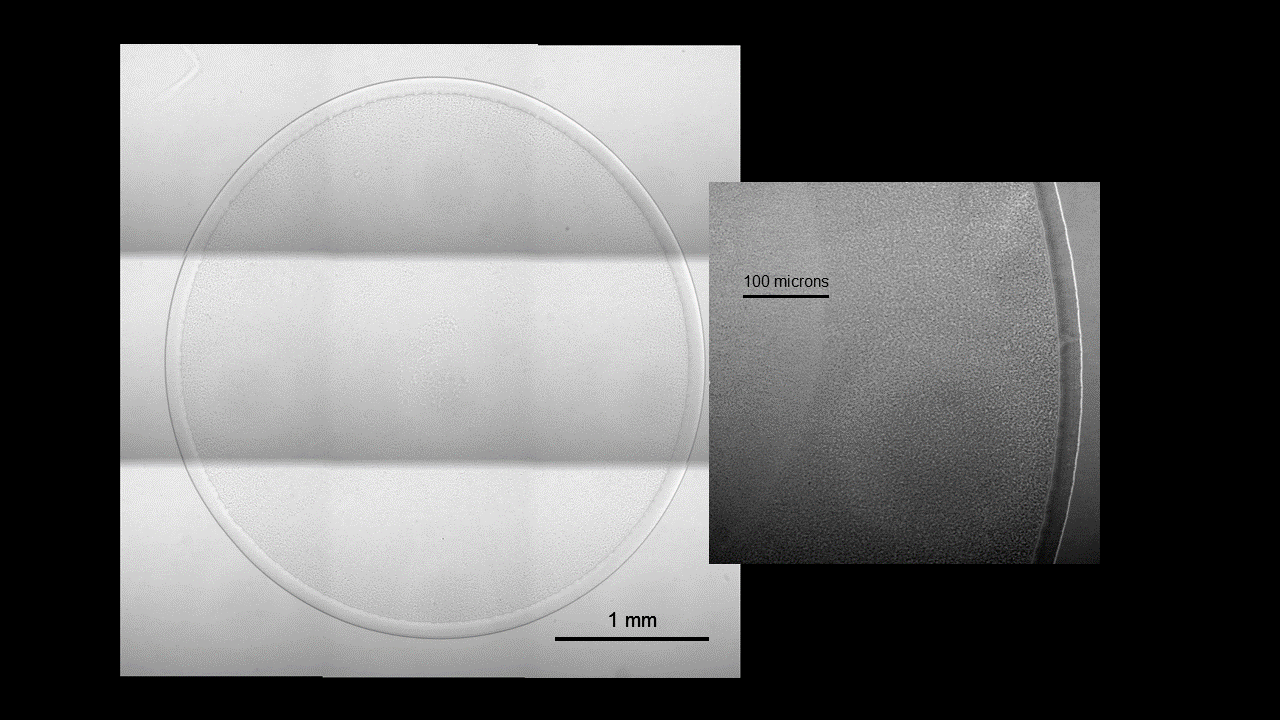}
\caption{\label{fig:bright_cR}Brightfield images showing the coffee-ring effect with \textit{V. Cholerae}}
\end{figure}


\begin{figure}
\centering

\begin{subfigure}[t]{.4\textwidth}
\centering
\includegraphics[width=\linewidth]{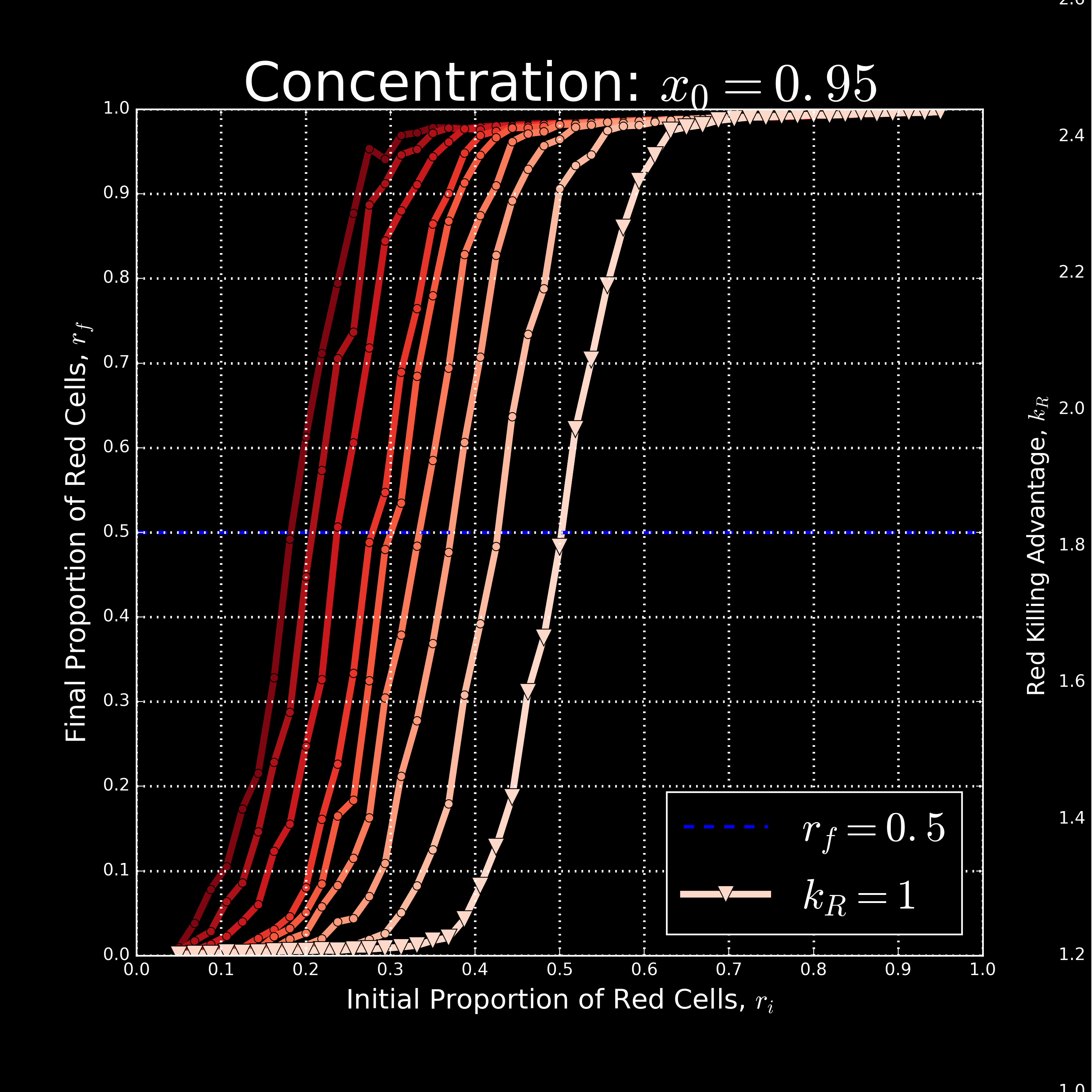}
        \caption{}\label{fig:fig_a}
\end{subfigure}
\begin{subfigure}[t]{.4\textwidth}
\centering
\includegraphics[width=\linewidth]{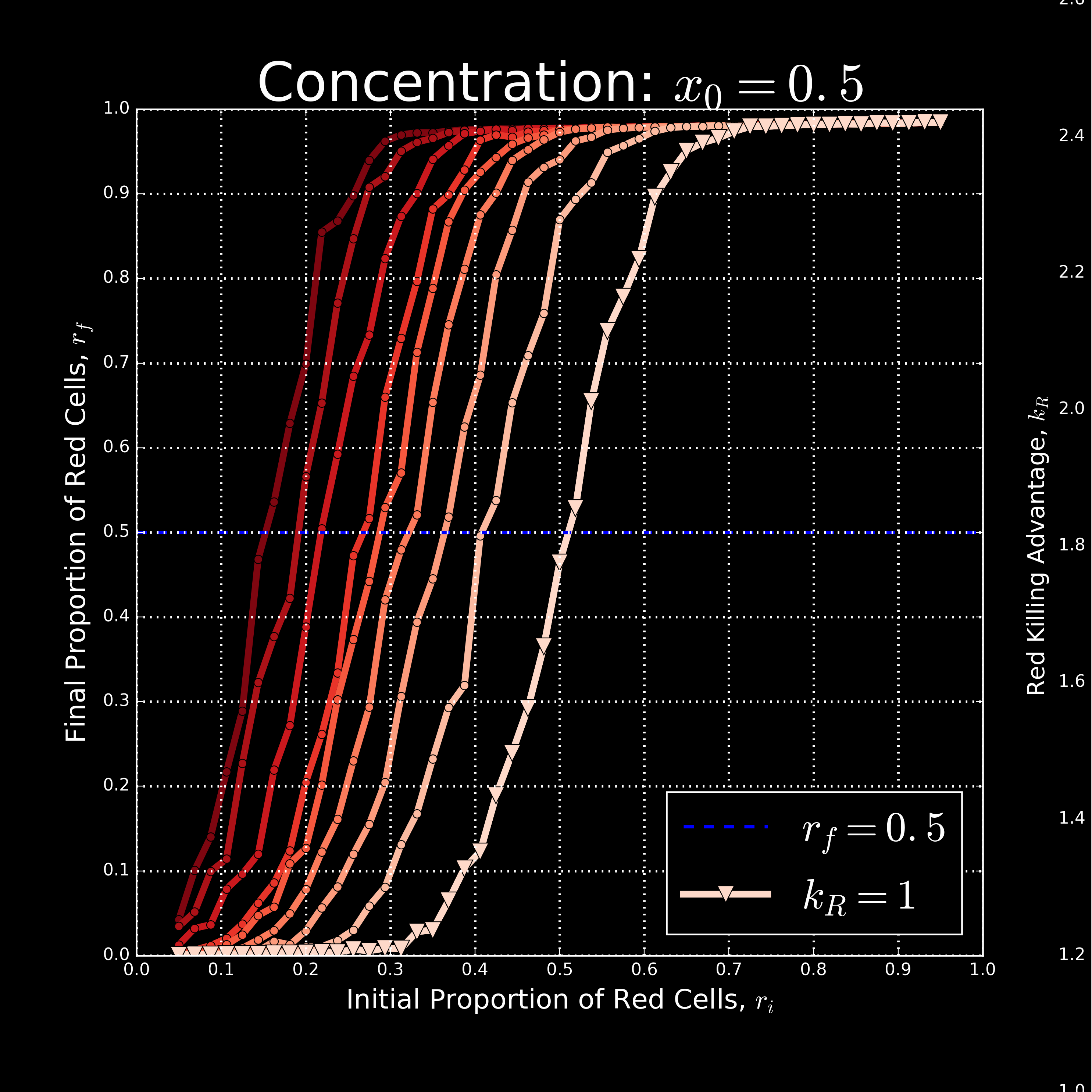}
\caption{}\label{fig:fig_b}
\end{subfigure}

\medskip

\begin{subfigure}[t]{.4\textwidth}
\centering
\includegraphics[width=\linewidth]{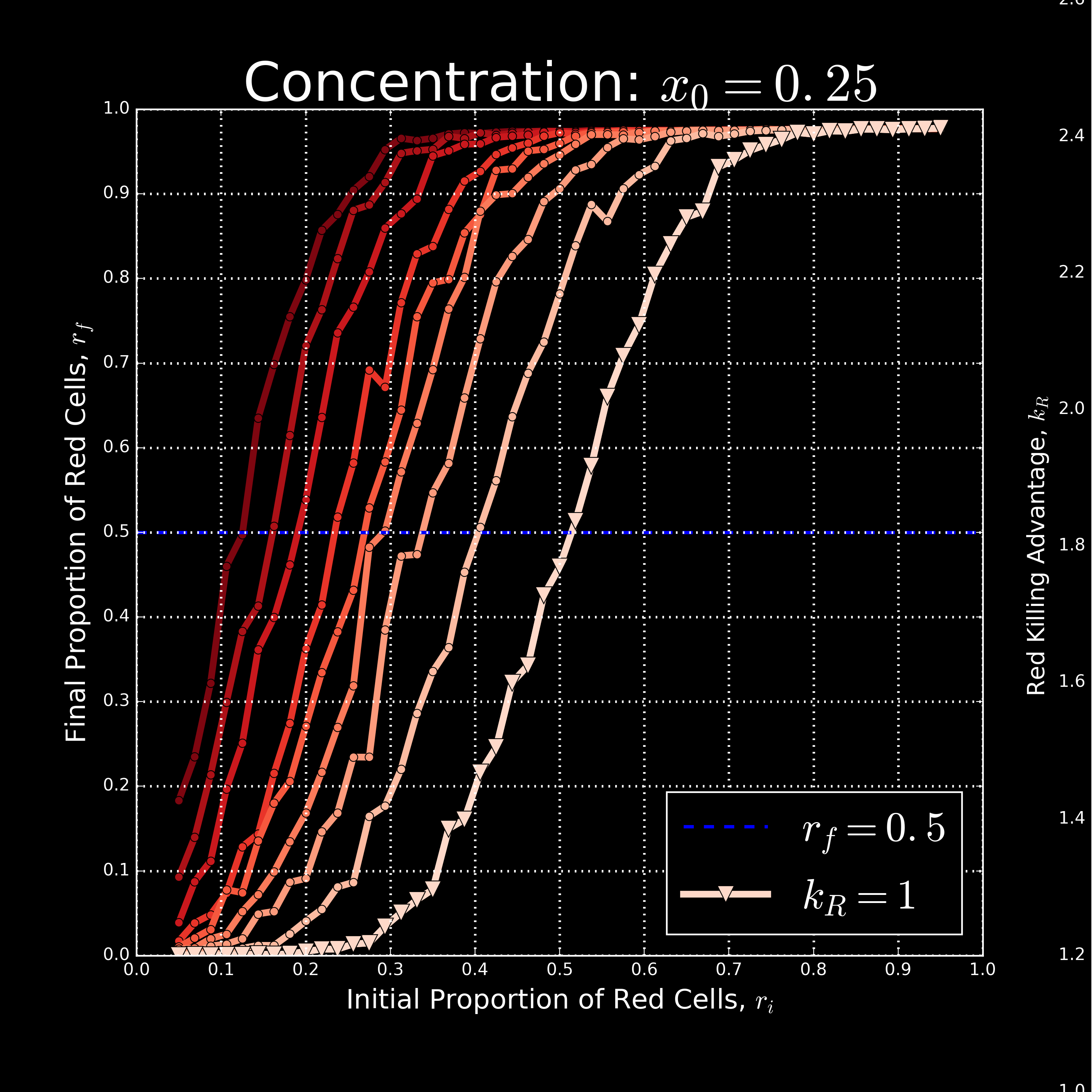}
	\caption{}\label{fig:fig_c}
\end{subfigure}
\begin{subfigure}[t]{.4\textwidth}
\centering
\includegraphics[width=\linewidth]{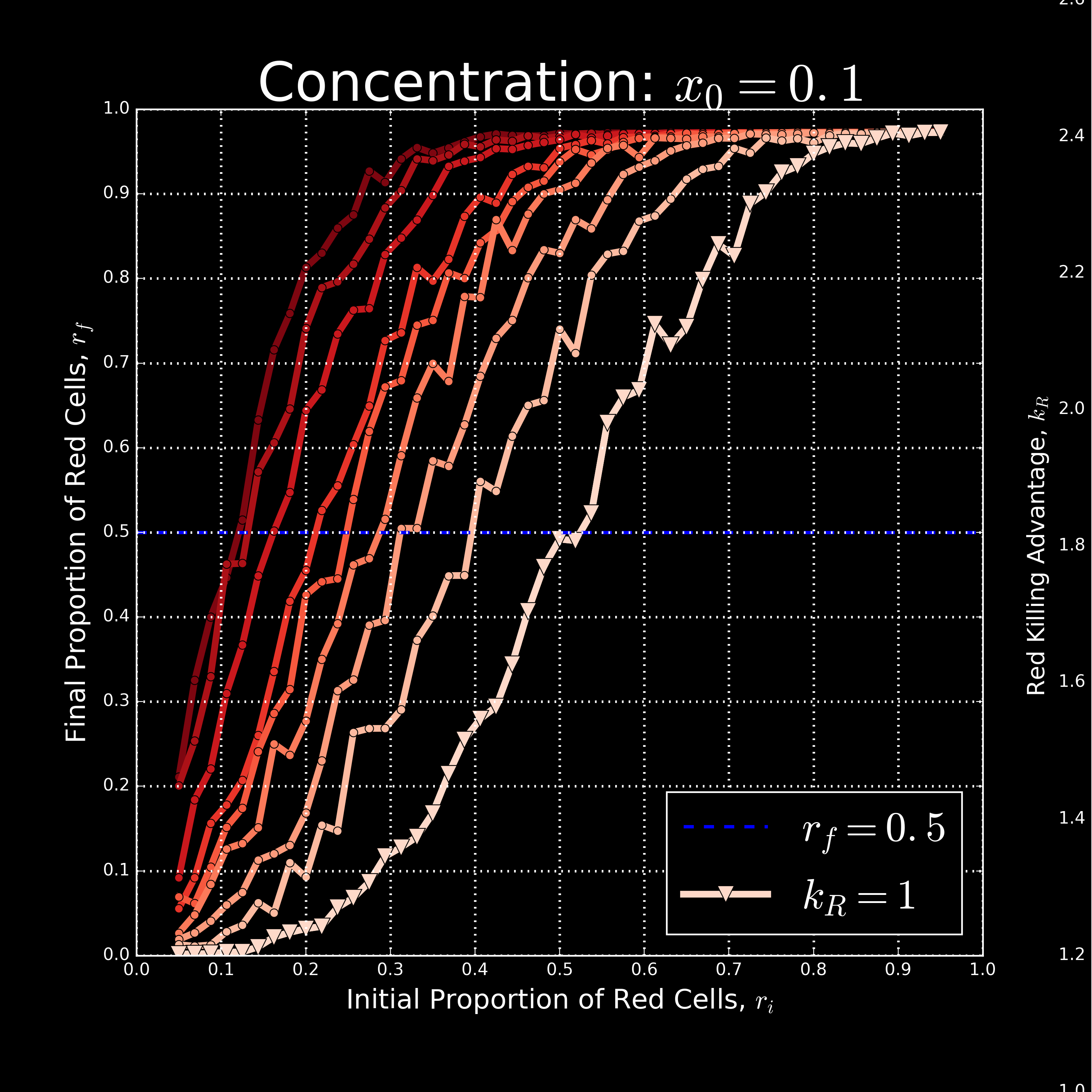}
\caption{}\label{fig:fig_d}
\end{subfigure}

\medskip

\begin{subfigure}[t]{.4\textwidth}
\centering
\includegraphics[width=\linewidth]{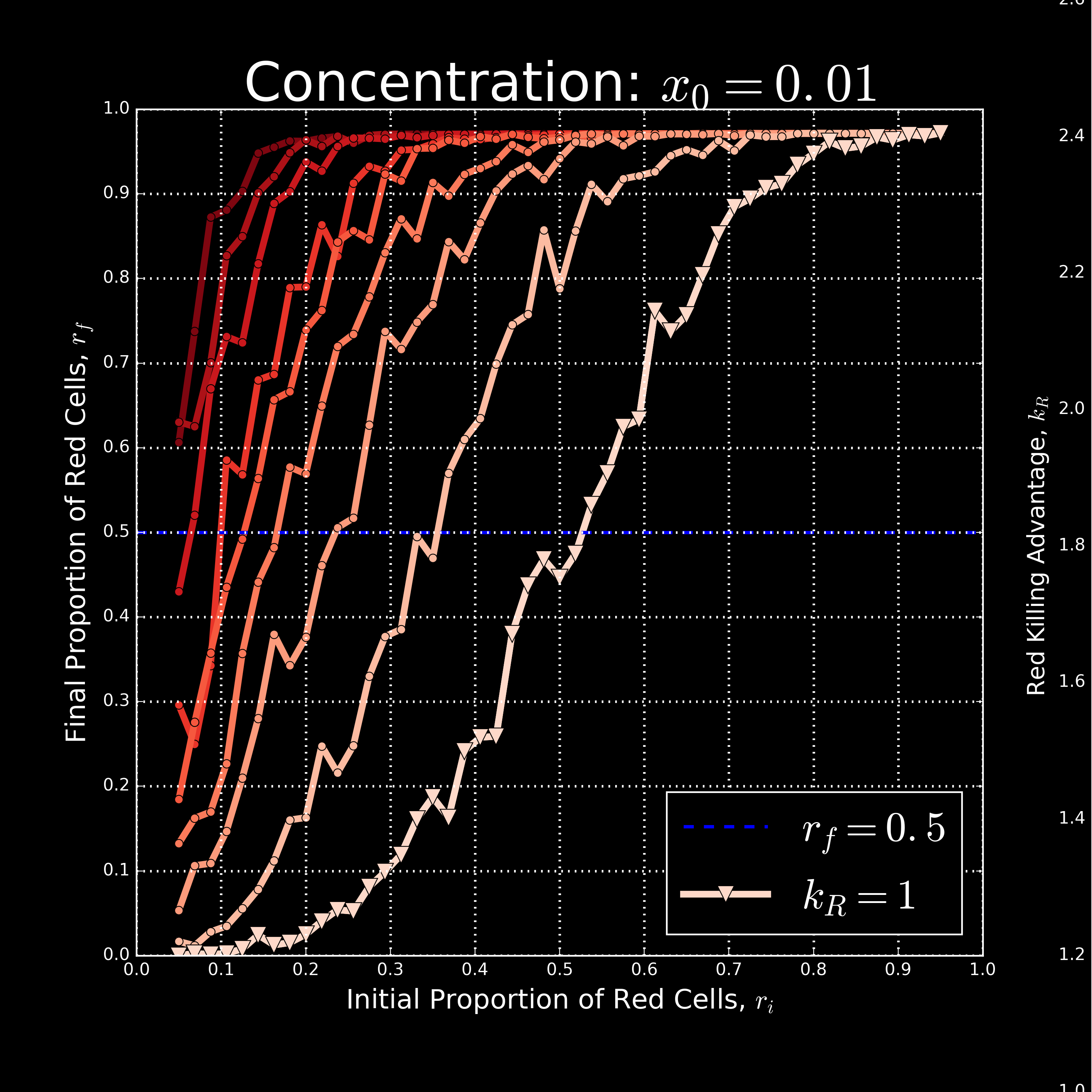}
\caption{}\label{fig:fig_e}
\end{subfigure}

\begin{minipage}[t]{\textwidth}
\caption{The final proportion of red cells, $r_{f}$, depends logistically on the initial proportion of red cells: $r_{f}=\frac{1}{1+e^{(-\beta(r_{i}-r_i^*))}}$.  (\subref{fig:fig_a}) shows the same data as main text figure 2A, while the other panels show families of logistic curves for simulations run at different initial cell concentration. $r_i^*$ describes the initial proportion of red cells required to "win" the competition, and can be located for a given curve by that curve's intercept with the blue line at $r_f=0.5$. As in the main text figure 2A, $k_R$ varies from $k_R=1$ (far right curve in each panel) to $k_R=2.7$ (far left curve in each panel). As initial concentration decreases, these curves shift left and appear noisier. The leftward shift is explained by the fact that the superior killer fares better in low initial density competitions, as discussed in the main text. The decrease in noise as density increases is due to the law of large numbers; the total number of cells becomes quite small as $x_0$ tends to zero, leaving the competition outcome susceptible to small initial fluctuations.}
\end{minipage}

\end{figure}

\begin{figure}[!ht]
\centering
\includegraphics[width=\textwidth]{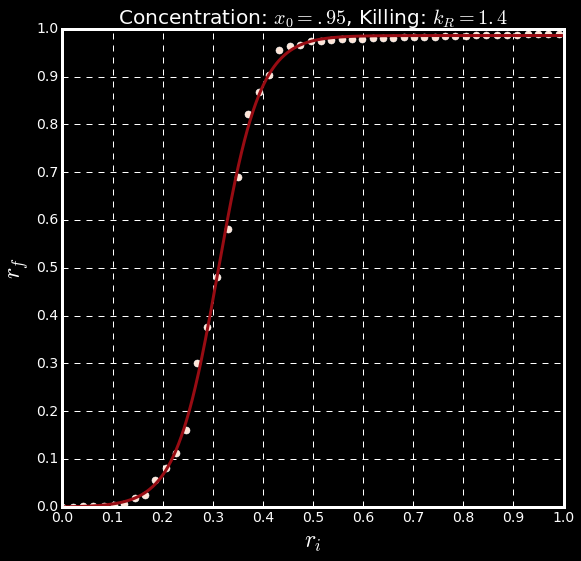}
\caption{\label{fig:sigmoid} To extract $r_i^*$ from simulations, we use SciPy's optimization module to fit data of the type shown in main text figure 2A and SI figure 5 to a sigmoid function: $r_{f}=\frac{1}{1+e^{(-\beta(r_{i}-r_i^*))}}$. This is a two free-parameter fit to fewer than 50 data points, which allows a sensible interpolation between simulation data points, affording an estimate of $r_i^*$ even when simulation data doesn't exist at exactly $(r_i=r_i^*,r_f=.5)$.}
\end{figure}

\begin{figure}[!ht]
\centering
\includegraphics[width=\textwidth]{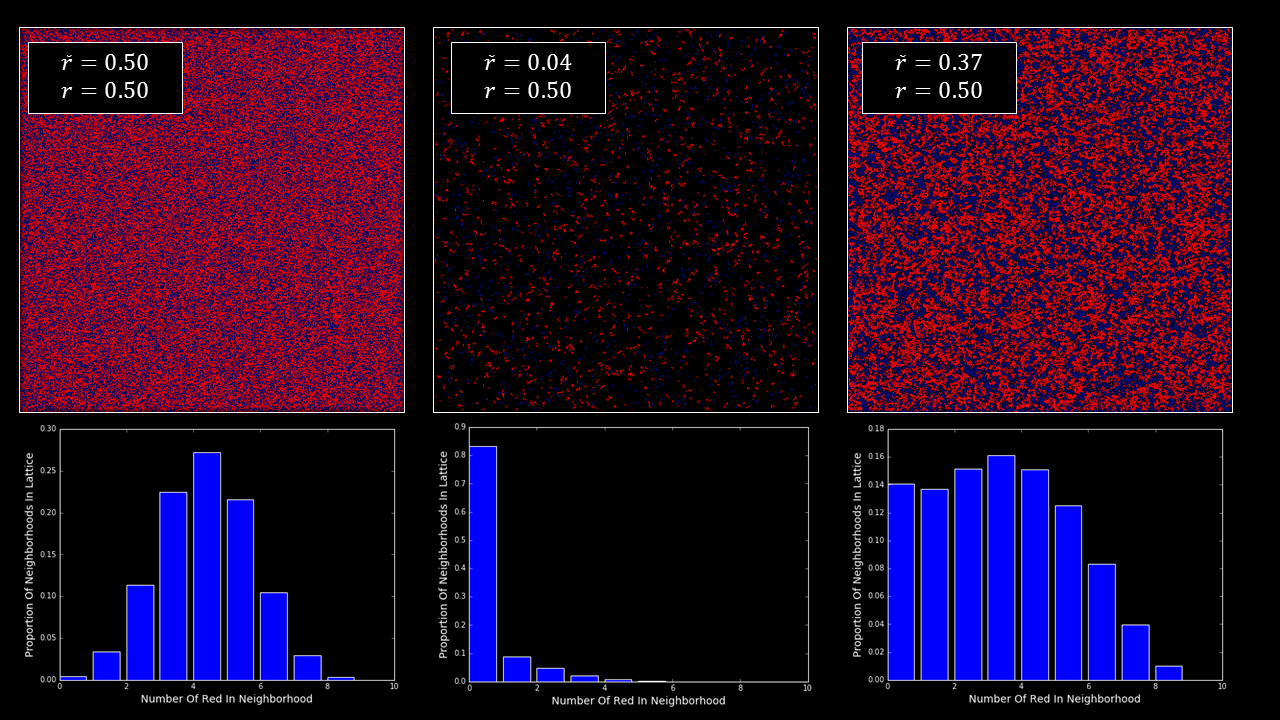}
\caption{\label{fig:r_tilde} Plotting $<\widetilde{r_i^*}>$ againts $k_R$ collapses the density dependence of competition outcome. Here the difference between $\widetilde{r}$ and $r$ is visualized. The three lattices shown all have equal total proportion of red to blue cells, but there is a difference in cellular arrangement which affects competition outcome. $\widetilde{r}$ captures this effect. The histograms below the lattices show the probability distribution for having $n$ red cells out of a $3\times 3$ neighborhood of $8$ cells (excluding the center cell). $<\widetilde{r}>$ is then calculated as $\frac{1}{8}\sum_{n=0}^8{n\cdot p(n)}$ }
\end{figure}

\end{document}